\documentclass[twocolumn,showpacs,10pt,aps,amssymb,amsmath,pre]{revtex4}

\usepackage{epsfig}

\newcommand{\jcop}{J. Comp. Phys.}

\newcommand{\be}{\begin{equation}}
\newcommand{\ee}{\end{equation}}
\newcommand{\bea}{\begin{eqnarray}}
\newcommand{\eea}{\end{eqnarray}}

\begin{document}

\title{On the fundamental diagram of traffic flow}

\author{Florian Siebel and Wolfram Mauser\\
        Department of Earth and Environmental Sciences \\ 
        University of Munich \\
        Luisenstra\ss e 37, D-80333 Munich, Germany
       }

\begin{abstract}
We present a new fluid-dynamical model of traffic flow. This model generalizes 
the model of Aw and Rascle [SIAM J. Appl. Math. 60 916-938] and Greenberg 
[SIAM J. Appl. Math 62 729-745] by prescribing a more general source term to 
the velocity equation. This source term can be physically motivated 
by experimental data, when taking into account relaxation and reaction time. 
In particular, the new model has a (linearly) unstable regime as observed 
in traffic dynamics. We develop a numerical code, which solves the
corresponding system of balance laws. Applying our code to a wide variety of
initial data, we find the observed inverse-$\lambda$ shape of the 
fundamental diagram of traffic flow. 
\end{abstract}
\date{\today}
\pacs{89.40.Bb, 05.10.-a, 47.20.Cq}
\maketitle

\section{Introduction}
After two-equation models of traffic flow were seriously criticized by 
Daganzo~\cite{Dag95} the main focus of the traffic community has shifted
towards microscopic models of traffic flow. However, the criticism has 
been overcome, see e.g.~\cite{Zha98,JWZ02}. By replacing the space derivate 
in old two-equation models by the convective derivative, Aw and 
Rascle~\cite{AwR00} and
Greenberg~\cite{Gre01} deduced a two-equation model, which solves all
inconsistencies of the earlier models as they showed with a detailed 
mathematical analysis and  numerical simulations. In particular, 
in their model (in the following called ARG model), no information travels 
faster than the vehicle velocity , i.e. in general drivers do not react to 
the traffic situation behind them. Moreover, the velocity is always 
non-negative. 
In the ARG model traffic flow is prescribed by the following system of balance 
laws determining the density $\rho = \rho(t,x)$ and velocity $v=v(t,x)$ of cars
\bea
\frac{\partial \rho}{\partial t} + \frac{\partial (\rho v)}{\partial x} & = & 0, \\
\frac{\partial (\rho(v-u(\rho))}{\partial t} + \frac {\partial (\rho v
  (v-u(\rho)))}{\partial x} & = & \frac{\rho
  (u(\rho)-v)}{T}.
\eea
As usual, $(t,x)$ denote the time and space variable. $u(\rho)$ denotes 
the {\it equilibrium velocity}, which fulfills the following conditions
\bea
\label{ucond1}
u'(\rho) & < & 0 \mbox{~for~} 0 \le \rho \le \rho_m, \\
\label{ucond2}
\frac{d^2 (\rho u(\rho))}{d \rho^2} & < & 0 \mbox{~for~} 0 \le \rho \le \rho_m,
\eea 
with the maximum vehicle density $\rho_m$. $T>0$ is an
additional parameter, the {\it relaxation time}. In the formal limit $T \to 0$ 
the ARG model reduces to the classic Lighthill-Whitham-Richards
model~\cite{LiW55,Ric56,Whi74}. For smooth solutions, the ARG model can be
rewritten as 
\bea
\label{rho}
\frac{\partial \rho}{\partial t} + v \frac{\partial \rho}{\partial x}  +
\rho \frac{\partial v}{\partial x}& = & 0, \\
\label{v}
\frac{\partial v}{\partial t} + (v + \rho u'(\rho)) \frac {\partial v}{\partial x} & = & \frac{u(\rho)-v}{T}.
\eea

In our opinion the ARG model still has a drawback, i.e. it can not explain the 
growth of structures and the general behavior for congested traffic, as 
observed in traffic dynamics (see e.g.~\cite{ScH04,Hel01}). To see this, we
consider a linear stability analysis around the equilibrium solution
$\rho(t,x) = \rho_0$, $v(t,x) = u(\rho_0)$, i.e.
\bea
\rho(t,x) & = & \rho_0 + \tilde{\rho} \exp(ikx + \omega(k)t), \\
v(t,x) & = & u(\rho_0) + \tilde{v} \exp(ikx + \omega(k)t).
\eea
Substituting this ansatz into system~(\ref{rho})-(\ref{v}) we obtain
\be
\begin{pmatrix}
\omega + i k u & i k \rho_0 \cr
-\frac{u'}{T} & \omega + \frac{1}{T} + i k (u + \rho_0 u')
\end{pmatrix}
\begin{pmatrix}
\tilde{\rho} \cr
\tilde{v}
\end{pmatrix}
= 
\begin{pmatrix}
0 \cr
0
\end{pmatrix}.
\ee
Nontrivial solutions of this linear system exist if and only if
\be
(\omega + i k u) \Big(\omega + \frac{1}{T} + i k (u + \rho_0 u')\Big) + i k \frac{\rho_0
  u'}{T} = 0,
\ee
or equivalently for
\bea
\omega_1 & = & - i k (u + \rho_0 u'), \\
\omega_2 & = & - \frac{1}{T} - i k u.
\eea
For the stability properties, the real parts of the above solutions are
important, i.e.
\bea
Re(\omega_1) & = & 0, \\
Re(\omega_2) & = & - \frac{1}{T}.
\eea
For $T>0$ both real parts are nonpositive, which means that the ARG
model is linearly stable, the velocity $v$ relaxes to the equilibrium velocity
$u$ in the entire region $0 \le \rho \le \rho_m$. This is clearly in contrast
to observations, where a wide range of states in the fundamental diagram, the
relation between vehicle flux and the density, are observed for congested
traffic flow. To cure this defect, Greenberg, Klar and Rascle developed an 
extended model with two equilibrium velocities~\cite{GKR03}. In 
the paper here, we propose an alternative model, which takes into account 
the reaction time of drivers (as well as mechanical restrictions).  

We give a physical argument for our new model and define it in
Sec.~\ref{derivation}. Section~\ref{implementation} presents the
methods used for numerically solving the model equations. 
Section~\ref{tests} describes tests to validate our numerical algorithm, 
before we finally discuss the numerical results on the fundamental diagram 
obtained with our model in Sec.~\ref{results}. 

\section{A heuristic derivation of the new model}
\label{derivation}
Before we turn to the new model, let us first give a simple derivation of
the ARG model. Note, that the model was mathematically derived from car 
following theory in~\cite{AKMR02}. Suppose that in the reference frame of 
individual drivers, drivers adjust their speed $v$ in such a way, that 
they asymptotically approach the equilibrium velocity $u$, i.e.
\be
\label{driver}
\frac{d(v-u)}{dt} = \frac{u-v}{T}.
\ee
Here, $T = \mbox{const} >0$ is the relaxation time. In comparison to optimal
velocity models (see e.g.~\cite{BHNSS95}) the equilibrium velocity term on the
left has been added which vanishes for $u = \mbox{const}$. It is easy to 
verify, that the analytical solution of the ordinary differential 
equation~(\ref{driver}) reads
\be
\label{drivere}
v(t) = u(t) + (v(0) - u(0)) \exp(-\frac{t}{T}).
\ee
In the coordinate system of the road, Eq.~(\ref{driver}) translates to
\be
\frac{\partial (v-u)}{\partial t} + v \frac{\partial (v-u)}{\partial x} = \frac{u-v}{T}.
\ee
Moreover, since
\be
- \Big( \frac{\partial u}{\partial t} + v \frac{\partial u}{\partial x} \Big)  
= - u' \Big( \frac{\partial \rho}{\partial t} + v \frac{\partial \rho}{\partial
  x} \Big)  =  \rho u' \frac{\partial v}{\partial x},
\ee
where we have used the continuity equation~(\ref{rho}) for the last equality, 
we recover the velocity equation of the ARG model~(\ref{v}). From this
derivation, it is obvious that drivers instantaneously react to the
current traffic situation.

We therefore tried to generalize Eq.~(\ref{driver}) and took the
reaction time of drivers $\tau > 0$ into account
\bea
\label{driverr}
\frac{dv}{dt}(x,t) - \frac{du}{dt}(\rho(x-v \tau,t-\tau)) = ~~~~~~~~~~~~~  \\
\nonumber
\frac{u(\rho(x-v \tau,t-\tau))  -  v(x-v\tau,t-\tau)}{T}. 
\eea
Using a Taylor series expansion in $\tau$ and keeping only terms up to order 0
in $\tau$ and $T$, i.e.
\bea
\nonumber
\frac{du}{dt}(\rho(x-v \tau,t-\tau)) & = & \frac{\partial u(\rho(x,t))}{\partial
  t} + v \frac{\partial u(\rho(x,t))}{\partial x} \\ 
&   & + O^1(\tau,T), \\
\nonumber
u(\rho(x-v \tau,t-\tau)) & = & u(\rho(x,t)) \\ 
\nonumber 
&   & - \tau u'(\rho(x,t))(
\frac{\partial \rho}{\partial t} + v \frac{\partial \rho}{\partial x}) \\ 
\nonumber
&   & + O^2(\tau,T) \\
& = & u + \rho u' \frac{\partial v}{\partial x} \tau + O^2(\tau,T), \\
\nonumber
v(x-v \tau,t-\tau) & = & v(x,t) - \tau \Big( \frac{\partial v}{\partial t} + v
\frac{\partial v}{\partial x} \Big) \\
&    & +  O^2(\tau,T),
\eea
we find
\be
\label{vi}
\frac{\partial v}{\partial t} + (v + \rho u'(\rho)) \frac {\partial
  v}{\partial x}  =  \frac{u(\rho)-v}{T-\tau}.
\ee 
This equation is identical to the velocity equation of the ARG
  model~(\ref{v}), except
that the relaxation time $T$ has been replaced by $T-\tau$. In particular it
follows from the stability analysis of the ARG model, that for $\tau >
T$ the new system is (linearly) unstable.

Before we look at the experimental data on the relaxation and reaction time,
we remark that one could be tempted to include an anticipation length into
the model, as e.g. in ~\cite{THH99,Nel00}. This approach has not been 
followed here for two reasons:
First, the ARG model already includes anticipatory elements, as noted
by~\cite{Gre01}. Second, including the anticipation length into the above
derivation yields a system, which does not guarantee that the maximum speed
at which information travels is bounded from above by the velocity of cars, 
and is therefore unrealistic.

\begin{figure}[htpb]
\includegraphics[width=\linewidth]{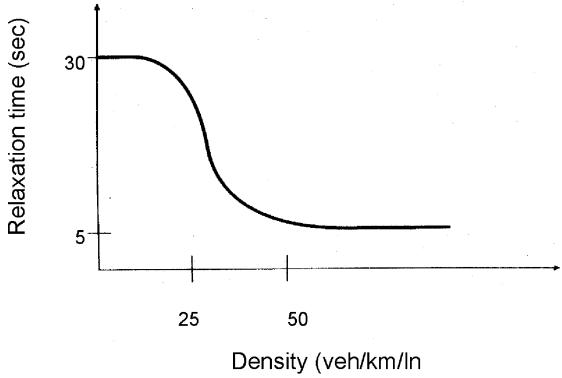}
\caption{Dependence of the relaxation time $\tilde{T}$ on the vehicle 
density per lane. Taken from~\cite{KuM97}.
\label{fig1}}
\end{figure}

For the reaction time $\tau$, typical values are of the order 
\be
\tau \approx 0.5~..~1~\mbox{s}.
\ee
Fig.~\ref{fig1} shows experimental results for the relaxation time 
$\tilde{T}$ taken from the review article~\cite{KuM97}.
However, these values have to be interpreted with care and cannot be translated
directly to our model context. To see this, we note that the relaxation time
$\tilde{T}$ is determined for the ansatz $v(t,x) = u(\rho(t+\tilde{T},x+\Delta
x))$, i.e. after the relaxation time $\tilde{T}$ the driver has fully adjusted
to the equilibrium velocity $u$. Here, according to Eqs.~(\ref{driver}) 
and~(\ref{drivere}), the equilibrium velocity in general will never be 
reached exactly. Instead, if we require, that $|v(t)-u(t)| <
|v(0)-u(0)|/1000$, we find that $t > 6.908~T \approx \tilde{T}$. I hence seems
reasonable to set
\be
\tilde{T} \approx 5~..~10~T.
\ee
Taking average values $\tau = 0.75~\mbox{s}$ and $\tilde{T} = 7.5~T$, we 
indeed find
that $T-\tau <0$ for about $\rho > 50$~[1/km/lane]. It was also pointed out
in~\cite{KuM97} that for large densities the relaxation
time $\tilde{T}$ increases, which means that $T-\tau > 0$ for $\rho \approx
\rho_m$.  

\begin{figure}[htpb]
\includegraphics[width=\linewidth]{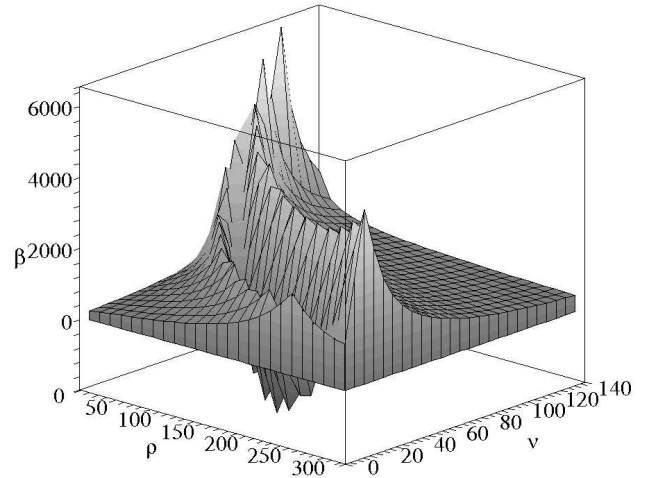}
\caption{Function $\beta(\rho,v)$ defined in Eq.~(\ref{beta}). We used units 
[$\rho$] = 1/km and [v]=km/h and [$\beta$] = 1/h.
\label{fig2}}
\end{figure}

One could try to repeat the derivation leading to Eq.~(\ref{vi}) for a
general relaxation time $T=T(\rho,v)$. Note, that the above derivation 
is only valid for a constant relaxation time. Moreover, it involves 
only the leading 
term of a Taylor series expansion. We therefore decided to generalize 
the velocity equation of the ARG model in the following way
\be
\label{vn}
\frac{\partial v}{\partial t} + (v + \rho u'(\rho)) \frac {\partial
  v}{\partial x}  =  \beta(\rho,v) (u(\rho)-v).
\ee
Note that we do not require $v \le u$ as Greenberg~\cite{Gre01}. From the
experimental data and the argument put forward before (note that the sign of
$\beta$ determines whether the traffic flow is linearly stable or not) we
require 
\bea
\label{beta1}
\beta(\rho,v) < 0 {\rm ~for~} 0 < \rho_1 < \rho < \rho_2 < \rho_m, v=u(\rho),\\ 
\label{beta2}
\lim_{\rho \to 0,\rho_m} \beta(\rho,v) > 0, \hspace{2cm} \\
\label{beta3}
\lim_{v \to 0,u_m=u(0)} \beta(\rho,v) > 0.  \hspace{2cm}
\eea
Throughout this paper we use a functional form
\be
\label{beta}
\beta(\rho,v)=  \left\{
\begin{array}{ll}
\frac{a_c}{u-v},& \mbox{if~} \tilde{\beta}(\rho,v)(u-v) - a_c \ge 0, \\
\frac{d_c}{u-v},& \mbox{if~} \tilde{\beta}(\rho,v)(u-v) - d_c \le 0, \\
\tilde{\beta}(\rho,v),& \mbox{else},
\end{array} \right.
\ee
where the function $\tilde{\beta}(\rho,v)$ is defined as
\be
\tilde{\beta} = \frac{1}{\hat{T}} \Big( 1 + \alpha \frac{|u-v|}{u_m} +
\frac{1}{\rho_1 \rho_2} (-(\rho_1 + \rho_2) \rho + \rho^2) \Big).
\ee
For the choice of the velocity-density relation of Cremer~\cite{Cre79}
\be
\label{Cremer}
u(\rho) = u_m \Big(1- \Big( \frac{\rho}{\rho_m} \Big)^{n_1}\Big)^{n_2}
\ee
and the parameters $\rho_m=300$~[1/km], $u_m=140$~km/h, $n_1 = 0.35$, $n_2=1$
(note that with these parameters, the equilibrium velocity of
Cremer~(\ref{Cremer}) fulfills the conditions~(\ref{ucond1}) 
and~(\ref{ucond2})), $\hat{T}=1$~s, $\alpha=12$, $\rho_1=70$~[1/km] and 
$\rho_2=270$~[1/km] the function $\tilde{\beta}(\rho,v)$ already fulfills all 
requirements~(\ref{beta1})-(\ref{beta3}). However, due to mechanical
restrictions, the maximum acceleration $a_c$ and deceleration $d_c$ give 
stronger limitations, i.e.
\be
\label{mechanicalrestriction}
\frac{dv}{dt} \le a_c \mbox{~and~} \frac{dv}{dt} \ge d_c
\ee
with typical values $a_c=2~{\rm m/s^2}$ and  $d_c=-5~{\rm m/s^2}$. Since the
resulting system is not strictly hyperbolic for equality in
Eq.~(\ref{mechanicalrestriction}), which is problematic for a numerical
solution, we prescribe the limitations on
\be
\label{mechanicalrestrictionn}
\frac{d(v-u)}{dt} \le a_c \mbox{~and~} \frac{d(v-u)}{dt} \ge d_c,
\ee
which then leads to the functional form~(\ref{beta}). We plot the function
$\beta(\rho,v)$ for the mentioned parameter values in Fig.~\ref{fig2}.
We stress that the above functions describe reality only qualitatively. For
realistic simulations of traffic flow, experimental data are required to
determine $\beta(\rho,v)$.

In the new model, traffic flow is described by the following system of
balance laws
\bea
\label{crho}
\frac{\partial \rho}{\partial t} + \frac{\partial (\rho v)}{\partial x} & = &
0, \\
\label{cv}
\frac{\partial (\rho(v-u(\rho))}{\partial t} + \frac {\partial (\rho v
  (v-u(\rho)))}{\partial x} & = & \beta  \rho (u - v),
\eea
or equivalently for smooth solutions by Eqs.~(\ref{rho}) and~(\ref{vn}). As the
corresponding system of the ARG model, the new system is strictly hyperbolic
for $0 < \rho \le \rho_m$.

\section{The numerical implementation}
\label{implementation}
Writing traffic flow as a system of balance laws in Eqs.~(\ref{crho})
and~(\ref{cv}) is very adequate for numerical purposes, as it
allows the application of well-established hydrodynamic methods for the 
numerical solution. We use a high-resolution shock-capturing scheme with
approximate Riemann solver for the numerical solution (see e.g.~\cite{LeV92}).

We rewrite Eqs.~(\ref{crho}) and~(\ref{cv}) in the form
\be
\frac{\partial U}{\partial t} + \frac{\partial F(U)}{\partial x} = S(U),
\ee
where
\bea
U & = & \begin{pmatrix} \rho \cr \rho (v-u) \end{pmatrix} =  \begin{pmatrix} U_1
  \cr U_2 \end{pmatrix}, \\
F(U) & = & \begin{pmatrix} \rho v \cr \rho v (v-u) \end{pmatrix} =
\begin{pmatrix} U_2 + U_1 u(U_1)
  \cr \frac{U_2^2}{U_1} + U_2 u(U_1) \end{pmatrix}, \\
S(U) & = & \begin{pmatrix} 0 \cr \beta \rho (u-v) \end{pmatrix}.
\eea
We use the second-order reconstruction scheme of van Leer~\cite{vLe77} to
reconstruct quantities at cell interfaces. At cell $i$ with cell center
at the location $x_i = x_0 + i \Delta x$ the update in time
from $t^n$ to $t^{n+1}$ is performed according to an conservative 
algorithm
\be
U_i^{n+1} = U_i^n - \frac{\Delta t}{\Delta x} (\hat{F}_{i+\frac{1}{2}} -
\hat{F}_{i-\frac{1}{2}}) + \Delta t S_i,
\ee
where $U_i^n = U(x_i,t^n)$ and $\Delta t = t^{n+1} - t^n$. To obtain a higher
order of convergence, we use the third order scheme of Shu and
Osher~\cite{ShO89}. The numerical fluxes $\hat{F}$ are determined according
to the flux-formula of Marquina~\cite{DoM96}, which reads
\be
\hat{F} = \frac{1}{2} (F^R+F^L-\Delta q).
\ee
Here, the superscripts $R$ and $L$ denote the reconstructed values on the 
right and left of a cell interface. The numerical viscosity term takes the form
\be
\Delta q = \mathbf{R}^R |\mathbf{\Lambda}|_{\max} \mathbf{L}^R U^R - \mathbf{R}^L |\mathbf{\Lambda}|_{max} \mathbf{L}^L U^L.
\ee
The matrix $|\mathbf{\Lambda}|_{\max}$ involves the characteristic speeds, 
\be
|\mathbf{\Lambda}|_{\max} = 
\begin{pmatrix}
\max(|\lambda_1^R|,|\lambda_1^L|) & 0 \cr
0 & \max(|\lambda_2^R|,|\lambda_2^L|)
\end{pmatrix},
\ee
where the characteristic speeds read explicitly
\bea
\lambda_1 & = & v + \rho u', \\
\lambda_2 & = & v.
\eea
$\mathbf{R}$ and $\mathbf{L}$ are the matrices of the right and left 
eigenvectors of the matrix
\be
\frac{\partial F}{\partial U} = 
\begin{pmatrix}
u + \rho u' & 1 \cr
- (v-u)^2 + \rho (v-u) u' & 2v-u
\end{pmatrix}
.
\ee
Explicitly,
\bea
\mathbf{R} & = & 
\begin{pmatrix}
1 & 1 \cr
u-v & v-u - \rho u'
\end{pmatrix},
\\
\mathbf{L} & = & \frac{1}{\rho u'}
\begin{pmatrix}
u-v+\rho u' & 1 \cr
v-u & -1
\end{pmatrix}.
\eea

\section{Code tests}
\label{tests}
We checked, that our numerical algorithm is convergent. Moreover, the density
equation~(\ref{crho}) is a strict conservation law. Prescribing periodic 
boundary conditions as in Sec.~\ref{results}, the total number of cars 
included in the numerical domain $\Omega$ should therefore be constant, i.e.
\be
\label{conservation}
\int_{\Omega} \rho ~dx = \mbox{const}.
\ee
We checked, that our numerical code fulfills Eq.~(\ref{conservation}) up to
machine precision (see also the corresponding results for a network
simulation based on the Lighthill-Whitham theory in~\cite{SiM05}).
\begin{figure}[htpb]
\vspace{0.8cm}
\includegraphics[width=\linewidth]{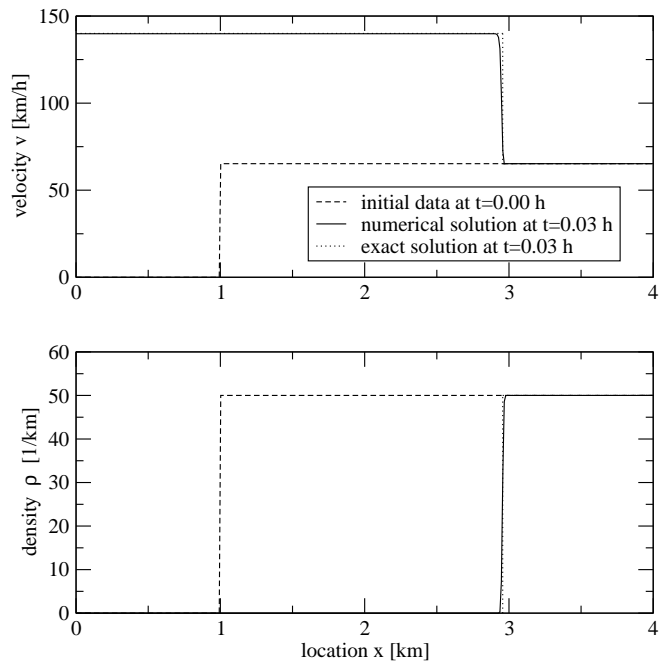}
\caption{Numerical solution for a Riemann problem of Aw and
  Rascle~\cite{AwR00}. The numerical solution at time t=0.03 h (solid line) 
for the initial data (dashed line) reproduces the exact solution (dotted
  line).  
\label{fig3}}
\end{figure}
\begin{figure}[t]
\hspace{1.0cm}
\includegraphics[width=\linewidth]{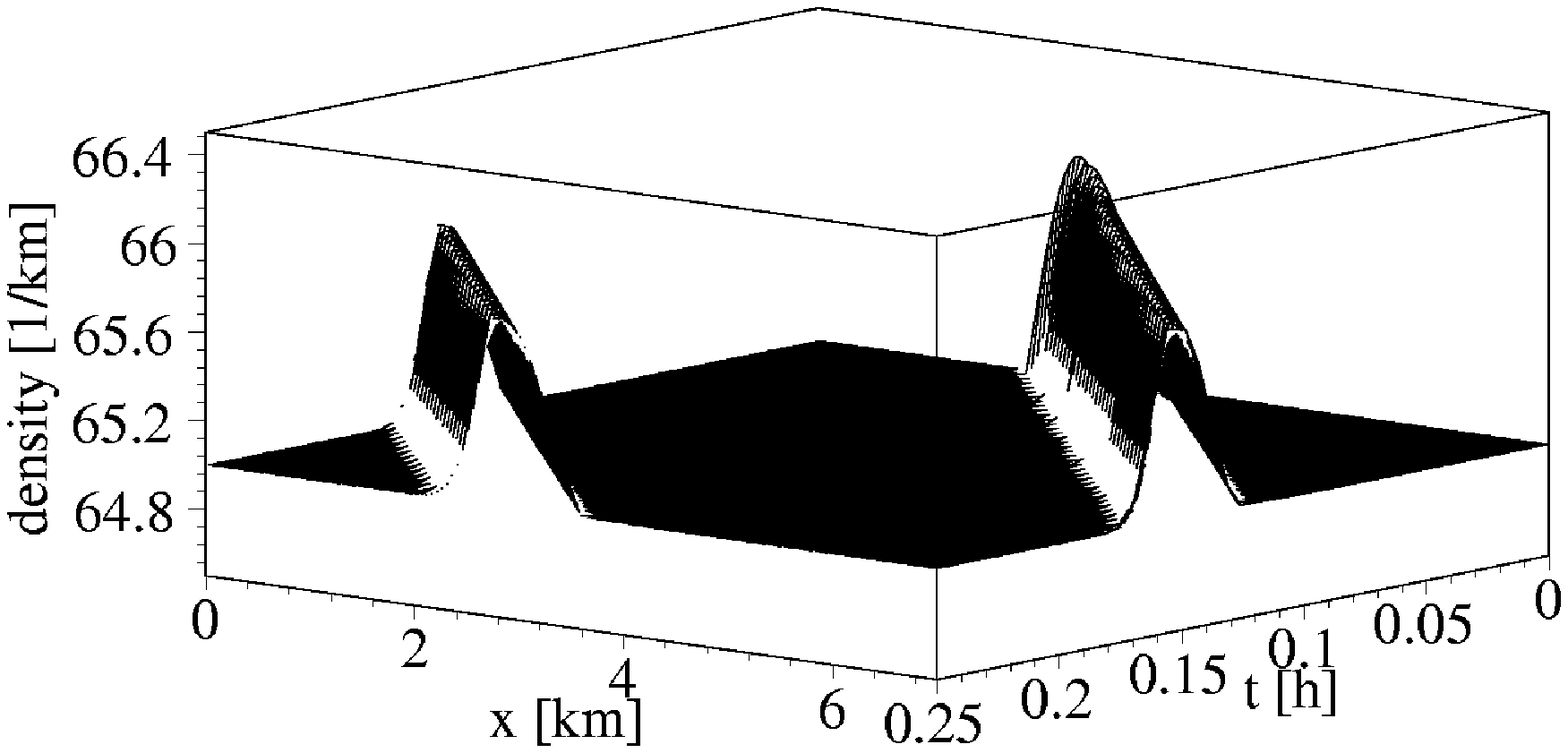}
\includegraphics[width=\linewidth]{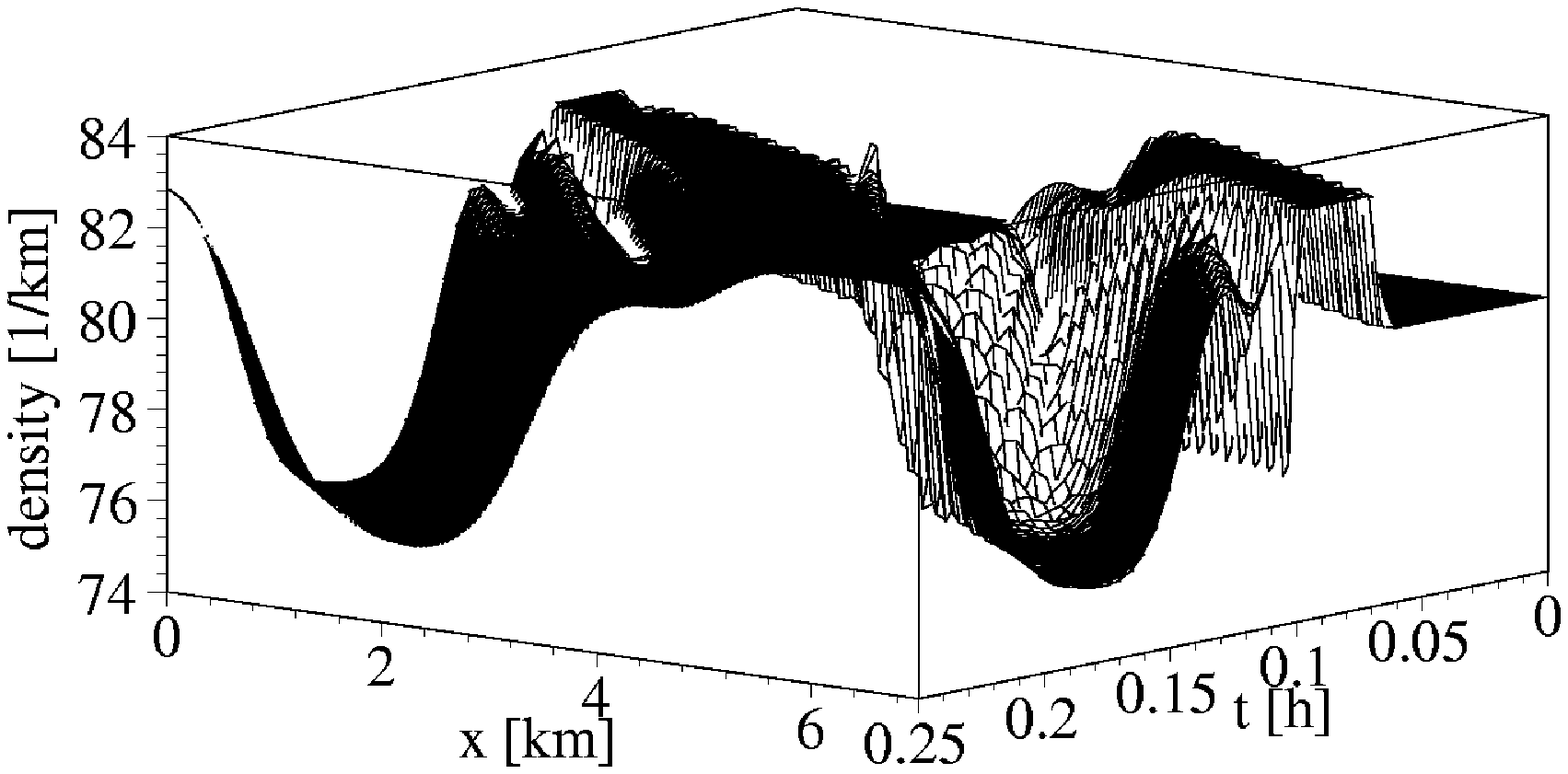}
\caption{Time evolution of the density for stable and instable initial data. 
We prescribe an equilibrium density $\rho_0 = 65$ [1/km] and $\rho_0 =
80$~[1/km] respectively, and on top a sinusoidal density
  perturbation. For the instable data, the
  initial perturbation located at $x=2.5$~[km] is amplified and finally two
  clusters form. 
\label{fig4}}
\end{figure}

Finally, Aw and Rascle presented in their paper the exact solution of a
Riemann problem, for which old two-equation models fail to describe the
correct behavior (see~\cite{AwR00}, Fig. 5.4). This Riemann problem consists
of the following initial data
\bea
\rho & = &  \left\{
\begin{array}{ll}
0,& \mbox{if~} x < 1 \mbox{~km}, \\
\rho_{+},& \mbox{if~} x \ge 1  \mbox{~km},
\end{array}
\right. \\
v & = &  \left\{
\begin{array}{ll}
0,& \mbox{if~} x < 1 \mbox{~km}, \\
v_{+},& \mbox{if~} x \ge 1 \mbox{~km}.
\end{array}
\right.
\eea
The exact solution to the homogeneous system consists of the constant state on
the right $(\rho_+,v_+)$ moving to the right with velocity $v_{+}$, leaving
behind vacuum. If we chose $v_+ = u(\rho_+)$, this exact solution will carry
over to our inhomogeneous system. Fig.~\ref{fig3} displays our
numerical solution for a choice $\rho_+ = 50$~[1/km]. For numerical reasons, we
prescribe a density $\rho = 10^{-6}$~[1/km] for $x<1$~km. Note, that our 
numerical algorithm resolves the steep gradient within only a few grid 
cells, at the same time reproducing the correct velocity at which the 
constant state moves to the right. Moreover, the velocity relaxes to the
equilibrium velocity behind the constant state $(\rho_+,v_+)$.

\section{Results on the fundamental diagram}
\label{results}
\begin{figure}[t]
\hspace{1.0cm}
\includegraphics[width=\linewidth]{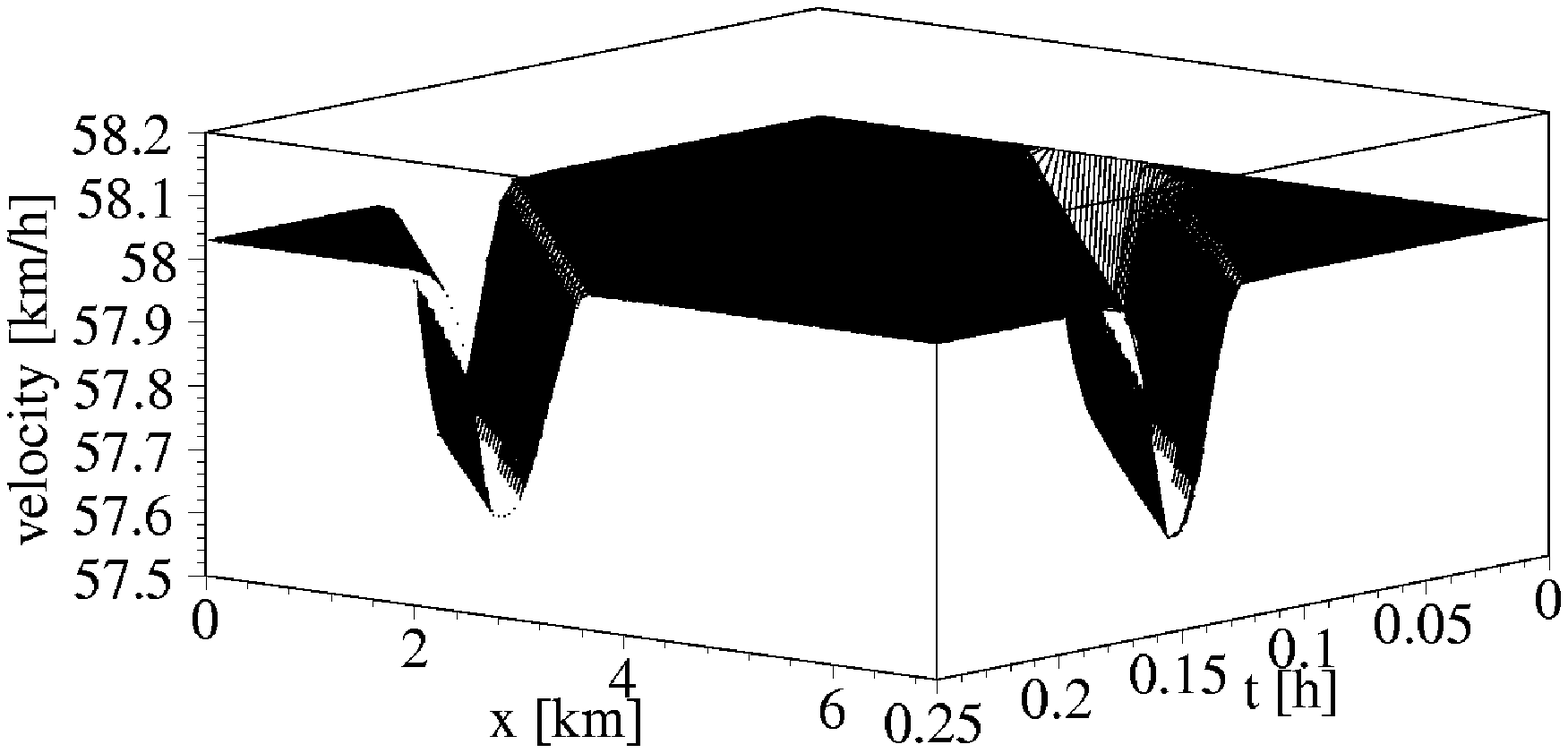}
\includegraphics[width=\linewidth]{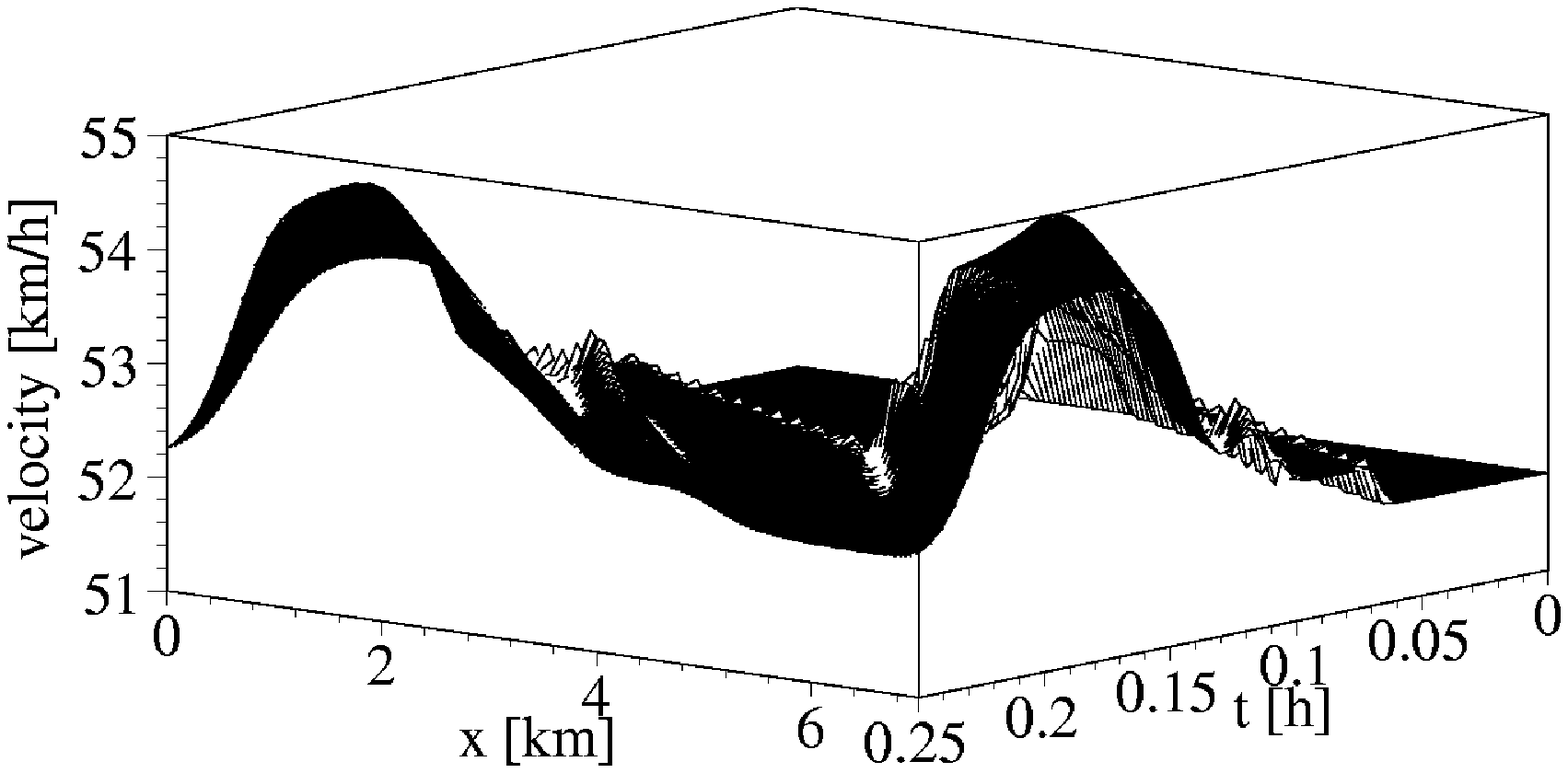}
\caption{Time evolution of the velocity for stable ($\rho_0 = 65$ [1/km]) and 
instable initial data ($\rho_0 = 80$~[1/km]) with initial density 
perturbation.  
\label{fig5}}
\end{figure}

For the results presented in this section we restrict us to a 7 km long
section of a (two-lane) highway with periodic boundary conditions. 
On this section of the highway, we start our simulations with
constant equilibrium data $\rho = \rho_0$, $v = u(\rho_0)$ and in addition 
between kilometers 2 and 3 a sinusoidal density perturbation
\be
\label{perturbation}
\Delta \rho = \sin(\pi x) {\ \rm~for~2~<~}x {\rm~<~3~km}.
\ee 
For all presented numerical results we used a resolution of 50 m.
Fig.~\ref{fig4} shows the evolution of these data for parameters $\rho_0 =
65$~[1/km] and $\rho_0 = 80$~[1/km]. Whereas the amplitude of the perturbation
is gradually damped 
with time for the stable initial data $\rho_0 = 65$~[1/km], the amplitude of
the perturbation increases for 
the instable initial data $\rho_0 = 80$~[1/km]. Moreover, the 
perturbation travels with a larger velocity downstream in the first case. 
For the unstable situation, two clusters are forming.
We plot the corresponding time evolutions of the velocity in Fig.~\ref{fig5}.

\begin{figure}[htpb]
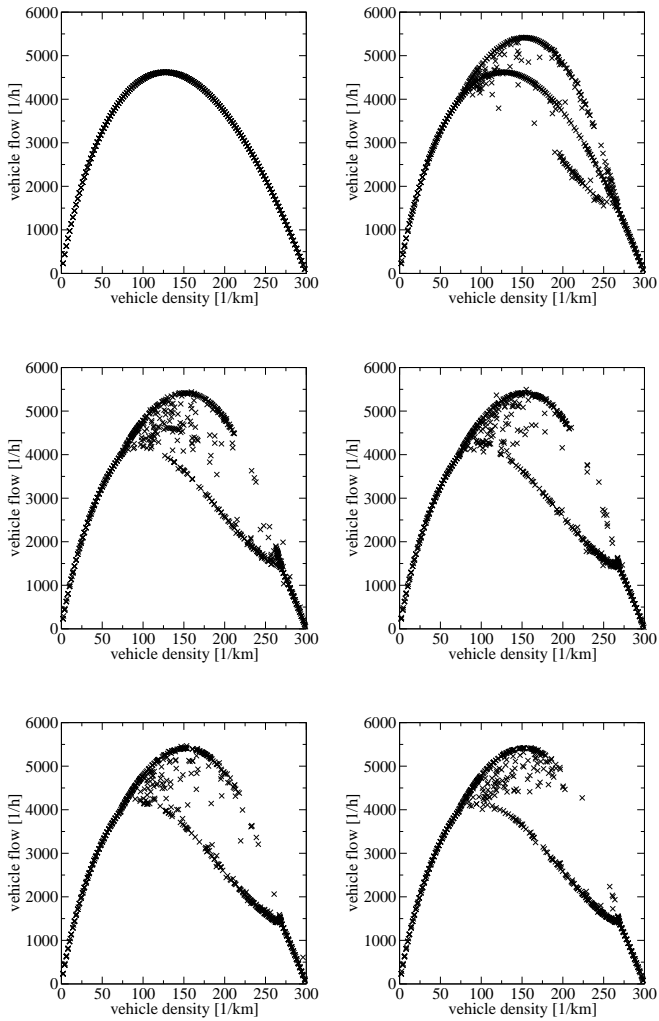

\vspace{0.3cm}
\noindent
\begin{minipage}[t]{.48\linewidth}
 \centering\epsfig{figure=fig6a.eps,width=\linewidth}
\vspace{0.3cm}
\end{minipage}\hfill
\begin{minipage}[t]{.48\linewidth}
 \centering\epsfig{figure=fig6b.eps,width=\linewidth}
\vspace{0.3cm}
\end{minipage}
\begin{minipage}[t]{.48\linewidth}
 \centering\epsfig{figure=fig6c.eps,width=\linewidth}
\vspace{0.3cm}
\end{minipage}\hfill
\begin{minipage}[t]{.48\linewidth}
 \centering\epsfig{figure=fig6d.eps,width=\linewidth}
\vspace{0.3cm}
\end{minipage}
\begin{minipage}[t]{.48\linewidth}
 \centering\epsfig{figure=fig6e.eps,width=\linewidth}
\vspace{0.1cm}
\end{minipage}\hfill
\begin{minipage}[t]{.48\linewidth}
 \centering\epsfig{figure=fig6f.eps,width=\linewidth}
\vspace{0.1cm}
\end{minipage}
 \caption{Fundamental diagram for the initial data (t = 0.00 h) (first row on
 the left), for t = 0.05 h (first row on the right), for t = 0.10 h (second
 row on the left), for  t = 0.15 h (second row on the right), for t = 0.20
 h (third row on the left) and for t = 0.25 h (third row on the right). At 
intermediate densities, our traffic model is linearly
 unstable, the representative points in the fundamental diagram are shifted 
towards two branches, which gives the visual impression of an
 inverted $\lambda$.\label{fig6}}
\end{figure}

To obtain a more general picture we varied the initial density in the entire
density regime and analyzed the resulting flow-density-relation as a function
of time. More precisely, we used initial values for the equilibrium data
$\rho_0 = 2,4,....,298$ and read off the resulting values for the density
$\rho$ and
the flux function $\rho v$ at 5 equidistantly distributed cross sections of
the highway. Fig.~\ref{fig6} shows the results for evolution times $t=0.00$~h
(initial data), $t=0.05$~h, $t=0.10$~h, $t=0.15$~h, $t=0.20$~h and $t=0.25$~h. 
For the
initial data, the flow-density curve corresponds closely to the equilibrium
flow density, the initial perturbation~(\ref{perturbation}) being negligible
for the visual output. After an evolution time $t=0.05$~h, the equilibrium
flow-density-curve is still visible, but in the unstable regime for densities
$70 < \rho < 270$ [1/km] two new flow-density-curves start to appear. In the
evolution further in time, the equilibrium density curve vanishes. Instead the
two new branches produce an inverse-$\lambda$ shape.

\section{Conclusion and Outlook}
\label{conclusion}
We generalized the traffic model of Aw, Rascle and Greenberg by prescribing a
more general source term to the velocity equation and developed a new
numerical code to solve the resulting system of balance laws. In total our 
(numerical) results show:
\begin{itemize}
\item
The new model can explain the large variance of the measured values of the
fundamental diagram in the congested regime, which correspond to fluctuations 
between two branches in the unstable density regime. Moreover, due to the 
stability properties, the model predicts oscillations in the relative 
velocity of cars in the congested regime, as they are found in experimental 
data. At the same time, it reproduces the small variance of velocities for 
free traffic flow and can explain the appearance of wide traffic jams.  
\item
Macroscopic traffic models have often used an equilibrium velocity $u(\rho)$, 
for which
$\frac{d(\rho u(\rho))}{d \rho^2} > 0 $ in the congested regime, in order to 
account for the values of traffic flow at the maximum (the tip of the 
inverted $\lambda$). According to our study, this is not necessary, as the 
high values for the fluxes can be explained with overcritical solutions 
and an equilibrium velocity function with 
$\frac{d(\rho u(\rho))}{d \rho^2} < 0$ everywhere.
\end{itemize}
The new model, which is a deterministic and effective one-lane model, has 
the capacity of reproducing many features
observed in traffic dynamics. In the presented work, the form of the function
$\beta$ in Fig.~\ref{fig2} was motivated by a physical argument, but the
quantitative details were determined rather ad hoc. However, we found that 
the fundamental diagram in the unstable region (e.g. the tip of the
inverted $\lambda$) depends on the particular form of $\beta$. Hence one should
try to determine the function $\beta$ from experimental data of the
fundamental diagram. In our opinion, the presented algorithm is
adequate for the use in network simulations of traffic flow.

\end{document}